\documentclass[aps,pre,twocolumn,superscriptaddress]{revtex4}
\usepackage{graphicx,subfigure}
\usepackage{amssymb}
\usepackage{amsmath}
\usepackage[british]{babel}
\usepackage{graphicx}

\begin{document}
%\preprint{preprint}
\title{Polymer Escape from a Metastable Kramers potential: Path Integral Hyperdynamics Study}
\author{Jaeoh Shin$^1$, Timo Ikonen$^2$, Mahendra D. Khandkar$^2$, Tapio Ala-Nissila,$^{2,3}$ and Wokyung Sung$^1$}
\affiliation{
$^1$Department of Physics, Pohang University of Science and Technology, Pohang 790-784, South Korea,\\ 
$^2$Department of Applied Physics and COMP Center of Excellence, Aalto University School of Science and Technology, P.O. Box 11100, FI-00076 Aalto, Espoo, Finland\\
$^3$Department of Physics, Brown University, Providence RI 02912-1843, U.S.A.}

\date{September 20th, 2010}

\begin{abstract}
  We study the dynamics of flexible, semiflexible, and self-avoiding polymer chains moving under a Kramers metastable potential. Due to thermal noise, the polymers, initially placed in the metastable
well, can cross the potential barrier, but these events are extremely rare if the barrier is much larger than thermal energy. To speed up the slow rate processes in computer simulations, we extend the recently proposed path integral hyperdynamics method to the cases of polymers. We consider the cases where the polymers' radii of gyration are comparable to the distance between the well bottom and the barrier top. We find that, for a flexible polymer, the crossing rate ($\mathcal{R}$) monotonically decreases with chain contour length ($L$), but with the magnitude much larger than the Kramers rate in the globular limit. For a semiflexible polymer, the crossing rate decreases with $L$ but becomes nearly constant for large $L$. For a fixed $L$, the crossing rate becomes maximum at an intermediate bending stiffness.  For a self-avoiding chain, the rate is a nonmonotonic function of $L$, first decreasing with $L$, and then, above certain length, increasing with $L$. These findings can be instrumental for efficient separation of biopolymers.

\end{abstract}

\maketitle

\section{Introduction}
The dynamics of polymer barrier crossing has attracted considerable attention in recent years, not only for basic understanding of numerous biological processes, but also for many practical applications. Biopolymers often need to surmount an entropic or energetic barrier for biological functions and in 
biotechnological applications, such as gene therapy, protein 
translocation, etc~\cite{Albert, Kasianowicz, SungPark}.
Han \textit{et al.}~\cite{Han} studied the
transport of double-stranded DNA (dsDNA) molecules through a
fabricated channel of alternating thickness, driven by electric
field. Within the relatively wide channels of one micron thick, the DNA are trapped in a coiled conformation. When an electric field drives the DNA into narrow channels, the chain, being severely confined and suffering a free energy barrier, becomes stretched. They found counterintuitively that the longer DNA molecules move faster than the shorter ones. The authors explained
it is attributable to the fact that longer DNAs have larger
contact area with the thin region and thus higher probability to escape the free
energy barrier. The origin of the free energy barrier is the competition
between the electric potential energy and the confinement entropy
characteristic of the polymer. A similar effect has been reported 
for polymer translocation through long nanoscale channels
\cite{Muthukumar03}.

This transport process can be viewed as an extension of the famous
Kramers problem~\cite{Kramers} of a Brownian particle crossing an
activation energy barrier. As an interconnected many-particle system, the long chain
polymer manifests cooperative dynamics in the presence of internal
noise and external fields. Depending upon various length scales
such as the chain contour length, radius of gyration, stretching and bending stiffnesses and other relevant parameters,
many features emerge. For contour length much smaller than the width
of a metastable potential as shown in  Fig.~\ref{fig:potential}, Park and Sung~\cite{ParkSung} developed
multi-dimension generalization of the Kramers
rate. They found that the rate of polymer
crossing the barrier is enhanced due to its flexibility. In
particular, the flexibility enables the chain coiled in the well
to stretch at the barrier top, which significantly lowers the activation energy, and enhances the barrier crossing rate. Also similar features were found in a double well potential, both for the flexible~\cite{SLee} and semiflexible ring polymers \cite{KLee}. In the opposite regime where the contour
length is much larger than the potential width,
Sebastian \textit{et al.}~\cite{Sebastian} suggested that the
flexible polymers cross the barrier by excitation and motion of a
kink and an anti-kink pair along the contour. 
Kraikivski \textit{et al.}~\cite{Kraikivski} studied a similar mechanism for
semiflexible polymers. Because the kinks are a local property of
the chain, activation energies are independent of the contour
length and the barrier crossing rate monotonically decreases with
the polymer contour length. For the intermediate cases of the semiflexible and self-avoiding
polymers where the radii of gyration are comparable to the width of the potential barrier, 
crossing dynamics has not yet been studied either analytically or
numerically.

Computer simulation is a useful tool to study complex systems,
complementing analytical methods. Since the rate of
polymer crossing over the barrier can become very small for
high barriers or long chains, conventional simulations
using either molecular dynamics or Brownian dynamics are
impractical because of long computing time required. To overcome this 
limitation, Voter~\cite{Voter} proposed the so-called hyperdynamics
(HD) simulation method of accelerating the rate by raising the well bottom with an appropriate 
correction factor for the cases where the transition state theory (TST) is valid. 
Recently, two groups (\cite{Chen}, \cite{nummela}) introduced the path integral hyperdynamics (PIHD) method, for the Langevin dynamics of a single Brownian particle.  
Unlike HD, PIHD allows an {\it exact} correction of accelerated 
dynamics without the TST assumption. 
In the present work we extend the PIHD method to 
a chain of many-particle systems, namely, polymers, where the internal degrees of freedom contribute significantly to the total free energy of the system.

We examine flexible, semiflexible, and self-avoiding polymers 
escaping a metastable well using the PIHD method. 
We consider the polymer's radius of gyration $R_g$ that is either
smaller or comparable to the width of the potential well. We focus on the
dependence of the escape rates on the contour length $L$ (or bead number $N$), chain stiffness, and excluded volume effect. We study how the chain's conformation and its transition affects the crossing rates.

The outline of the paper is as follows. In the next section, we
first recapitulate the path integral hyperdynamics method  
for the case of a single particle and then extend the method to polymer chains.
In Sec.~\ref{sec:models}, we describe our simulation models and methods, 
whose results are discussed in Sec.~\ref{sec:results}. Finally, we conclude and summarize our results in Sec.~\ref{sec:conclusions}.

\section{The path integral hyperdynamics method}
\label{sec:PIHD}

\subsection{The single-particle case}

A Brownian particle moving subject to a potential $V(\vec{r})$ is described by the Langevin equation,
\begin{eqnarray}
 m\ddot{\vec{r}}(t)+\zeta\ \dot{\vec{r}}(t)+ \nabla V(\vec{r}) =\vec{\xi }(t),
\end{eqnarray}
where $m$ is the mass, $\zeta$ is the friction coefficient 
and $\vec{\xi}$ is Gaussian white random force with  
$\langle \vec{\xi}(t)\rangle =0$ and 
$\langle \xi_{p}(t)\xi_{q}(0)\rangle=2\zeta k_{B} T\delta_{p,q}\delta (t) $. 
Here, $\langle \cdots \rangle$ denotes the ensemble average, $p$ and $q$ are cartesian coordinate indices, $k_{B}$ is the Boltzmann constant, and $T$ is the absolute temperature. Hereafter we drop the vector notation for brevity.
The probability density of finding the
particle at $r_{f}$ at time $t$ given an initial position $r_{0}$ at time $t_{0}$ is
\begin{eqnarray}
P(r_{0}, t_{0}|r_{f}, t)= C\int[Dr]\exp \lbrace-\beta I[r(t)]\rbrace,
\end{eqnarray}
where $C$ is a normalization constant, $\beta = (k_{B}T)^{-1}$, and $[Dr]$ represents the path integral over all possible trajectories $r(t)$, and the effective action is given by
\begin{eqnarray}
I[r(t)]=\frac{1}{4\zeta}\int_{t_{0}}^{t} dt'[m \ddot{r }(t')+\zeta\ \dot{r}(t')+ \nabla V(r) ]^2.
\end{eqnarray}

In a system with an energy barrier much larger than the thermal 
energy $k_\mathrm{B}T$, the probability of the particle crossing 
the barrier is very small. To make such transition events more 
frequent, a bias potential $V_\mathrm{bias}(r)$ is 
added to the actual potential $V(r)$.  In the boosted potential, 
$V_\mathrm{b}(r) \equiv V(r)+V_\mathrm{bias}(r)$, the particle obeys the Langevin equation
\begin{eqnarray}
 m \ddot r(t)+\zeta\ \dot{r}(t)+ \nabla V_\mathrm{b}(r) =\xi (t).
\end{eqnarray}
Obviously, this leads to dynamics and transition probabilities that
are different from the those given by Eqs. (1) and (2). However,
as shown by \cite{Chen} and \cite{nummela}, 
it is possible to exactly recover the original probability density of Eq.~(2)
from the biased dynamics by writing
\begin{eqnarray}
P(r_{0}, t_{0}|r_{f}, t)= C\int[Dr]\exp (-\beta I_\mathrm{b}[r(t)])\exp (-\beta I_{\xi}[r(t)]),
\end{eqnarray}
where the effective action can now be written in two parts: 
the action in the boosted potential ($I_\mathrm{b}$) 
and the correction factor
\begin{eqnarray}
I_{\xi}(t)=\frac{1}{4\zeta} \int_{t_{0}}^{t}dt'\nabla V_\mathrm{bias}(r(t'))[\nabla V_\mathrm{bias}(r(t'))-2\xi(t')].
\end{eqnarray}

To calculate the transition rates from the transition probability density, 
it is convenient to use the transition path sampling method~\cite{Chandler}. 
Here, the sampling is done over all dynamical paths $r(t)$ 
starting from a pre-transition state $1$ ($x_{1}<x_{c}$) 
at time $t=t_{0}$ to state $2$ located at $x_{2}>x_{c}$ at time $t$, where
 $x_{c}$ represents a certain transition state. The phenomenological 
 rate constant $\mathcal{R}$ is then given by the relation  
$\mathcal{R}(t)=dP_{1\rightarrow 2}/dt= \mathcal{R}\exp(-t/t_{r})$, where 
$P_{1\rightarrow 2}(t)$ is the transition probability
\begin{eqnarray}
P_{1\rightarrow 2}(t)=\int_{x_{f}\geq x_{c}} dr_{f} \int_{x_{0}\leq x_{c}} dr_{0}P(r_{0})P(r_{0},t_{0}|r_{f},t),
\end{eqnarray}
and $t_{r}$ is the transition time. 
The first and the second integrals are calculated over all 
accessible post-transition and all pre-transition states given by the 
initial quasiequilibrium distribution $P(r_{0})$ of the particle. 
For barriers larger than thermal energy, the rate reaches a well 
defined plateau after an initial transient period. 
Sampling all the events that have started with the initial 
configurations and crossed the transition state 
\textit{under the boosted potential,} $P_{1\rightarrow 2}(t)$ is given by
\begin{eqnarray}
P_{1\rightarrow 2}(t)=\frac{1}{n} \sum_{\xi} \exp(-\beta I_{\xi}(t)), \label{eq:crossing_prob}
\end{eqnarray}
where $n$ is number of all the paths and the 
summation is over the crossing paths only.

\subsection{Extension to polymer chains with internal degrees of freedom}

To date, the PIHD method has been demonstrated to work for systems where
there are no internal degrees of freedom \cite{Chen,Khandkar09}. It is
clear from the PIHD formalism that the external bias potential affects the
evolution of these internal degrees of freedom in a many-particle system and its
entropy changes. Thus, any transition rates that are influenced by an entropic
contribution to the free energy barrier are not necessarily correctly described
by the formalism. 

In this section we extend the PIHD method for systems with internal degrees of
freedom. In particular, we consider the case of polymer chains consisting of
$N$ beads and interacting with each other via a potential $U$.
The position $r_{i}$ of the $i$th bead of the polymer can be 
described by the Langevin equation
\begin{eqnarray}
m \ddot r_{i}(t)+\zeta\ \dot{r_{i}}(t)+ \nabla_{i} \Phi(r_{i}) =\xi_{i}(t),
\end{eqnarray}
where $\Phi(r_{i})=V(r_{i})+U$.

For the center-of-mass (CM) coordinate, $R(t)=(1/N)\sum_{i} r_{i}(t)$, which we regard as the reaction coordinate of the barrier crossing dynamics, we have
\begin{eqnarray}
 M \ddot{R }(t)+N\zeta \dot{R}(t)+ \sum_{i}\nabla_{i} V(r_{i}) = \Xi(t),
\end{eqnarray}
where $M$ is total mass ($Nm$) and $\Xi(t)=\sum_{i}\xi_{i} (t)$ is a Gaussian random force that satisfies $\langle\Xi(t)\rangle=0$ and $ \langle\Xi_{p}(t)\Xi_{q}(0)\rangle=2N\zeta k_{B} T\delta_{p,q}\delta (t) $. Applying bias potentials to all beads, we have
\begin{eqnarray}
 M \ddot{R }(t)+N\zeta \dot{R}(t)+ \sum_{i}\nabla_{i} V(r_{i})+\sum_{i}\nabla_{i} V_\mathrm{bias}(r_{i}) =\Xi (t).
\end{eqnarray}

We calculate the transition rate in the same way as in the single-particle case. However, now we define that the final state is reached whenever the polymer's center of mass has crossed the potential barrier. The probability is given by Eq.~(\ref{eq:crossing_prob}), except that now the correction factor is
\begin{eqnarray}
I_{\Xi}(t)=\frac{1}{4N\zeta} \int_{t_{0}}^{t}dt'\sum_{i}\nabla_{i}V_\mathrm{bias}(r_{i}(t')) [\sum_{i}\nabla_{i}V_\mathrm{bias}(r_{i}(t'))-2\Xi(t')].
\end{eqnarray}
%
%\begin{eqnarray}
%P(R_{1}, t_{1}|R_{2}, t_{2})= \exp\{-\frac{I_{\Xi}(t)}{k_{B}T}\}\int[DR]P[\Xi(t)]\delta(R_{a}(t_{2})-R_{2})
%\end{eqnarray}

In the polymer barrier crossing problem, the choice of bias potential is a critical one. Because the polymer conformation has a significant effect on the crossing rate through its entropic contribution to the free energy barrier, in particular for long chains~\cite{ParkSung, SLee, KLee}, the bias potential should be chosen in such a way that it does not affect the polymer conformation. With a properly chosen bias, only the energetic part of the barrier is changed, while the entropic part remains unchanged.

In the absence of bias potential, the conformational free energy of the chain with the CM position given as $R$ is
\begin{eqnarray}
F_{0}(R)=-k_{B}T\ln Z_{0}(R)\label{eq:free_energy},
\end{eqnarray}
 where
\begin{eqnarray}
Z_{0}(R)=\int \prod_{i}dr_{i}\delta(\frac{1}{N}\sum_{i} r_{i}-R)\exp(-\beta (\sum_{i}V(r_{i})+U)).
\end{eqnarray}
Here $\int \prod_{i} dr_{i}$ stands for integration over all possible internal configurations of the polymer. When we add a bias potential $V_\mathrm{bias}(r_{i})$ to each polymer segment, the free energy is given as
\begin{eqnarray}F(R)=-k_{B}T\ln Z(R),
\end{eqnarray}
where
\begin{eqnarray}Z(R)=\int \prod_{i}dr_{i}\delta(\frac{1}{N}\sum_{i} r_{i}-R)\exp(-\beta (\sum_{i}V(r_{i})+U)-\beta \sum_{i}V_\mathrm{bias}(r_{i})).
\end{eqnarray}

For an arbitrary choice of bias $V_\mathrm{bias}(r_{i})$, $F(R)-F_{0}(R)\neq \sum_{i}V_\mathrm{bias}(r_{i})$, which means that adding the bias changes the entropic part of the activation barrier. However, if we choose a uniform force bias $V_\mathrm{bias}(r)=-br$, we have
\begin{eqnarray}
F(R)-F_{0}(R)= \sum_{i}V_\mathrm{bias}(r_{i}).
\end{eqnarray}
That is, this particular choice of the bias potential does not change the entropy or the polymer's conformation. By choosing a uniform force bias, the crossing probability is given by Eq.~(\ref{eq:crossing_prob}), with the correction factor $I_{\Xi}(t)$ determined by
\begin{eqnarray}
I_{\Xi}(t)=\frac{b}{4\zeta} \int_{t_{0}}^{t}dt'(bN+2\Xi(t')).
\end{eqnarray}
We note that in Ref.~\cite{nummela} the PIHD method was used to study pulled biopolymers without considering the influence of bias to the free energy.

\section{Polymer models and simulation methods}
\label{sec:models}

In this paper, we consider three different polymer models. In all cases, the polymers are modeled as bead-spring chains, with the interaction potential $U$ chosen differently for the three models.  In the simplest model of a flexible polymer, we include the stretching energy only, {\it i.e.}, $U=U_s$, where
\begin{eqnarray}
U_{s}=\sum_{i=1}^{N-1} \frac{1}{2} k\left(|r_{i}-r_{i+1}|-l_{0}\right)^2.
\end{eqnarray}
Here $l_{0}$ is the natural bond length and $k$ is the spring constant. In the second case, we also include the bending energy such that $U=U_{s}+U_{b}$. Here
\begin{eqnarray}
 U_{b} = \sum_{i=2}^{N-1} \frac{1}{2} \kappa \left(r_{i-1}-2r_{i}+r_{i+1}\right)^2,
\end{eqnarray}
where $\kappa$ is the bending stiffness.

As the third model, we consider the self-avoiding FENE chain, with the interaction potential given by the finite extension nonlinear elastic (FENE) potential and the short-range repulsive Lennard-Jones (LJ) potential: $U=U_\mathrm{F}+U_\mathrm{LJ}$. The FENE potential is defined between neighboring monomers as
\begin{equation}
U_\mathrm{F}=-\sum_{i=1}^{N-1}\frac{1}{2}k_\mathrm{F}R_0^2\ln\left(1-\left(r_{i}-r_{i+1}\right)^2/R_0^2\right),
\end{equation}
where $R_0$ is the maximum allowed separation between connected monomers. The LJ potential is defined as
\begin{equation}
U_\mathrm{LJ}=\sum_{i<j}^{N}4\epsilon\left[\left(\sigma/r_{ij}\right)^{12} -\left(\sigma/r_{ij} \right)^6\right],
\end{equation}
for $r_{ij}\leq 2^{1/6}\sigma$ and 0 for $r_{ij} > 2^{1/6}\sigma$. Here $r_{ij}=|r_{i}-r_{j}|$ is the separation of the monomers, $\sigma$ is the diameter of the monomer and $\epsilon$ is the depth of the potential.

\begin{figure}
\includegraphics[width=8.4cm] {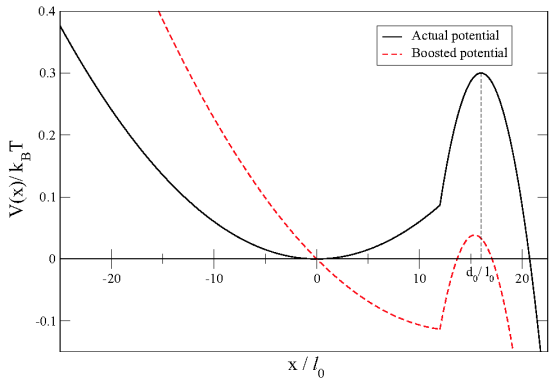}
\caption[0]{Actual potential $V(x)$ (solid line) and the boosted potential $V_{b}(x)$ (dashed line) we study.\label{fig:potential}}
\end{figure}
We consider a two-dimensional space with an external metastable potential that depends on $x$ as in Fig.~\ref{fig:potential} while $y$ represents the coordinate lateral to the potential force. The external potential is a one-dimensional piecewise-harmonic potential, defined by the equations
\begin{eqnarray}
V(x)&=& \frac{1}{2} \omega_{0}^2 x^2~\mathrm{for}~x< x_{0};\label{eq:extpot1}\\
V(x)&=& V_{B}-\frac{1}{2} \omega_{B}^2 (x-d_{0})^2~\mathrm{for}~x>x_{0},\label{eq:extpot2}
\end{eqnarray}
where $\omega_{0}^2$ and $\omega_{B}^2$ are the curvatures at the potential well and barrier, respectively, and $V_{B}$ is the potential barrier energy per segment. The position of the barrier is $d_{0}$ and $x_{0}$ is the crossover point between the piecewise harmonic potential.

The position of each monomer as a function of time is given by the Langevin equation
\begin{eqnarray}
m \ddot r_{i}(t)+\zeta\ \dot{r_{i}}(t)+ \nabla_{i} \left[ \Phi(r_{i})-br_i \right] =\xi_{i}(t),\label{eq:single_langevin}
\end{eqnarray}
where $-br_i$ is the bias potential. The Langevin equations~(\ref{eq:single_langevin}) are integrated in time by the method described by Ermak and Buckholz~\cite{Ermak, Allen}. Initially, the system is equilibrated without the bias potential, i.e., $b=0$. Then the bias is switched on to expedite barrier crossing. The transition state of the chain crossing is $X=X_{c}$, where $X_{c}$ is CM position where CM free energy $F(X)$ is the maximum. Due to asymmetry of the potential $V(x)$, the $X_{c}$ is different from $d_{0}$~\cite{Sebastian06}, and found to be smaller than $d_{0}$. Therefore we can choose $X=d_{0}+2l_{0}$ ($X=d_{0}+2\sigma$ for the FENE chain) such that, whenever CM reach these positions the crossing occurs with very few chains recrossing back to the well.

For the first two cases, we use the parameters $l_{0}$, $m_{0}$ and $(k_{B}T/1.2)$ to fix the length, mass and energy scales, respectively. We consider one bead as three bases of dsDNA, for which $l_{0}=1.02$ nm and $m_{0}\approx 1870~\mathrm{amu}$ and the characteristic time is $ t_{0}=\sqrt{\frac{1.2m_{0}l_{0}^2}{k_{B}T}}=30.9$ ps. For the FENE chain, we fix the parameters $\sigma$, $m$ and $\epsilon$. The time scale is then given by $t_\mathrm{LJ}=\left(m\sigma^2/\epsilon \right)^{1/2}$. The dimensionless parameters in all of our simulations are $m =1$, $\zeta=0.7$, $k_{B}T=1.2$ and, in addition for the FENE chain, $k_\mathrm{F}=15$ and $R_0=2$. For the spring constant $k$ and bending stiffness $\kappa$ we use various values, as indicated in Sec.~\ref{sec:results}. For a given $\kappa$ the persistence length is $2\kappa/k_{B}T$ in 2D.

The dimensionless curvature of potential
well and barrier are set to be $\frac{13}{9} \times 10^{-3}$ and
$3.2\times 10^{-2}$, which correspond to $1.157\times 10^{-3}
k_{B}T/{\rm nm}^2$ and $2.56\times 10^{-2} k_{B}T/{\rm nm}^2$,  respectively.
We choose $V_{B}=0.3 k_{B}T$, $x_{0}=12l_{0}$ ($x_0=12\sigma$ for the FENE chain) and
$d_{0}=16l_{0}$ ($d_0=16\sigma$).

\section{Results and Discussion}
\label{sec:results}

\subsection{Flexible polymer chains}

\begin{figure}
\includegraphics[width=8.5cm] {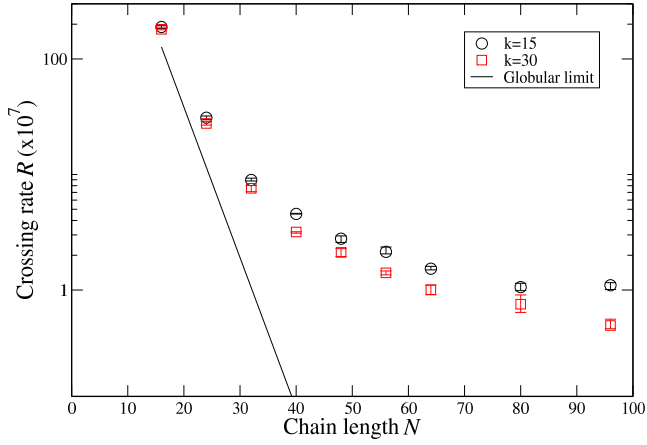}
\caption[0]{Rate of flexible polymer barrier crossing for $k=15$, $k=30$ and the globular limit ($\mathcal{R}_0$) as a function of the chain length $N$. \label{fig:rate_flexible}}
\end{figure}

 For a flexible polymer, we first study its crossing rate as a function of polymer length $N$ for two values of spring constant $k=15,~30$.  Figure~\ref{fig:rate_flexible} shows that the crossing rate decreases with $N$, but is still much larger than the Kramers rate~\cite{Kramers} $\mathcal{R}_{0}=\omega_{0}\omega_{B}/(2\pi\zeta)\exp(-\beta NV_{B})$ that the polymer has in the globular limit ($k\rightarrow \infty, l_{0}\rightarrow0$). The enhancement of the rates over this limit, larger for smaller $k$, is due to the chain flexibility that induces an entropy increase and conformational change in surpassing the potential barrier~\cite{ParkSung, SLee}. 

\begin{figure}
\includegraphics[width=8.5cm] {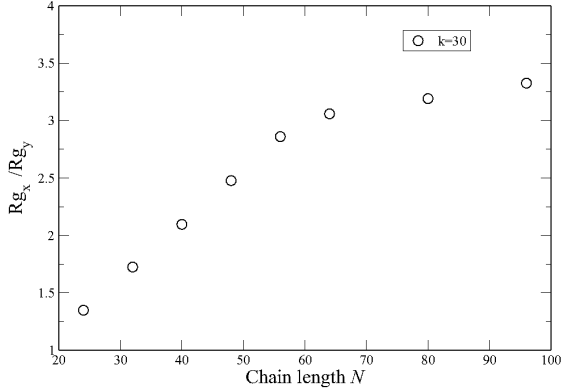}
\caption[0]{The ratio of the radii of gyration along $x$ and $y$ axis for the flexible chain ($k=30$) at the barrier top.} \label{fig:Radius}
\end{figure}

We find that as the chain becomes longer, its configuration at the barrier top (i.e., with the center of mass $X_{CM}$ placed at $d_{0}$) changes from a coiled state to a stretched state. Figure~\ref{fig:Radius} shows a dramatic increase of $R_{g,x}/R_{g,y}$ with $N$, where $R_{g,x}^2\equiv \langle \sum_{i}^{N} (x_{i}-X_{CM})^2 \rangle /N$ and $R_{g,y}^2\equiv \langle \sum_{i}^{N} (y_{i}-Y_{CM})^2 \rangle /N$ are the radii of gyration along $x$ and $y$ axes respectively at the barrier top, leading to further enhancement of the rate $\mathcal{R}$ over the globular limit $\mathcal{R}_{0}$.

\begin{figure}
\includegraphics[width=8.5cm]{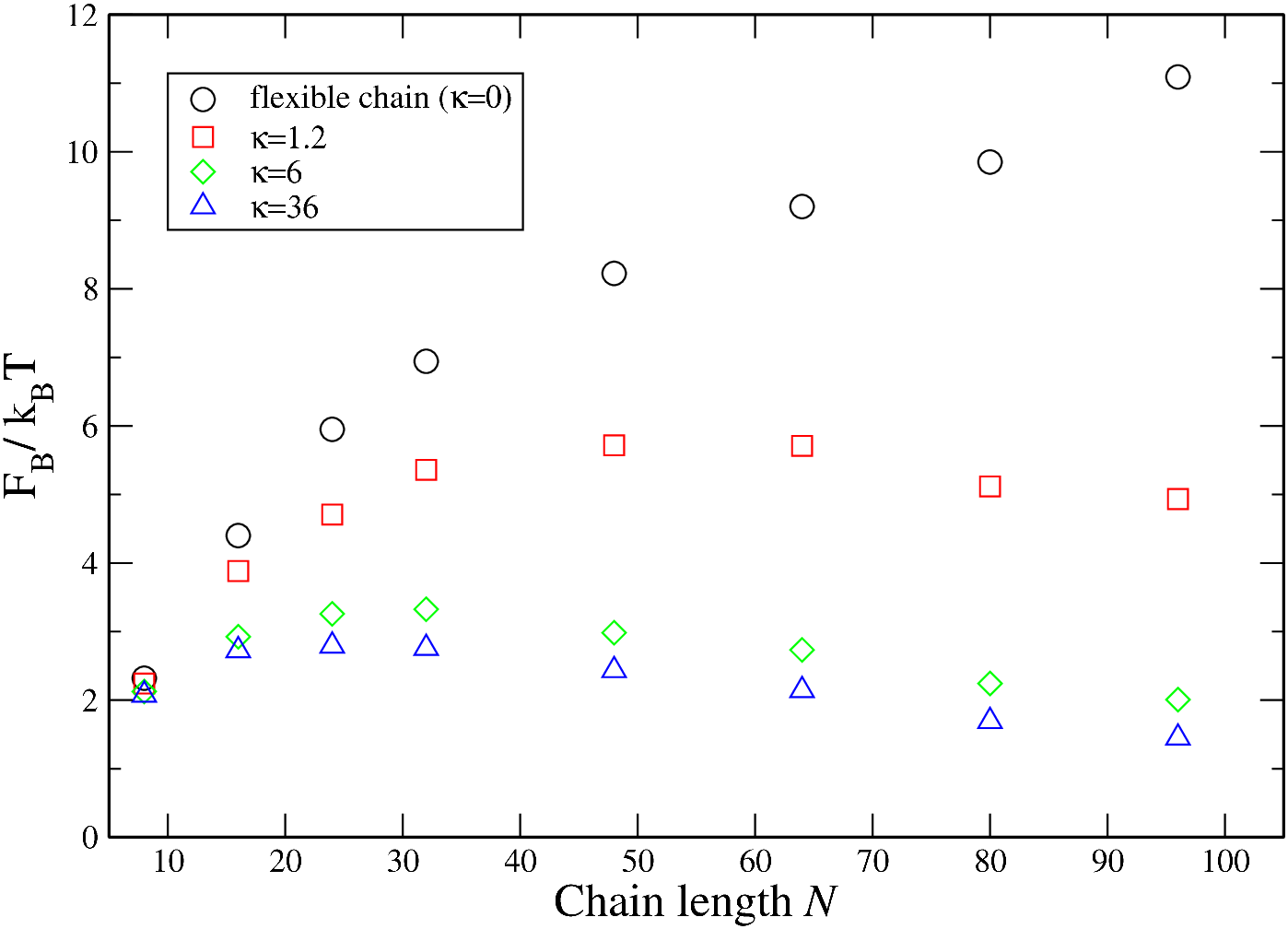}
\caption[0]{The free energy barrier height $F_{B}$ as a function of chain length for a flexible chain with $k=30$ and semiflexible chain with different bending stiffnesses $\kappa$. While $F_{B}$ increases monotonically for the flexible chain, it has a turnover behavior for the semiflexible chains. \label{fig:height}}
\end{figure}

 Provided that the escape dynamics is much slower than the segmental relaxations, one can consider the dynamics as that of CM in a free energy $F(R)$ [Eq.~(\ref{eq:free_energy})]. We have obtained the free energy by averaging over all configurations at a fixed CM position. The activation free energy barrier height $F_{B}$, the free energy difference between well bottom and barrier top, are shown in Fig.~\ref{fig:height}. In contrast to Ref.~\cite{ParkSung}, where $F_{B}$ for a stretched conformation can be reduced dramatically by a factor proportional to $N^3$, it monotonically increases with $N$. This is consistent with an analytical study of Sebastian and Debnath done for a similar system~\cite{Sebastian06}. In the present case, a long chain stretches partly threading not only around the barrier top but also around the well bottom. This conformation does not significantly reduce $F_{B}$ and enhance the crossing rate as in Ref.~\cite{ParkSung}.

On the other hand, for a case of $L(=Nl_{0})$ much larger than $d_{0}$, it is found that the activation energy is independent of polymer length, so that the barrier crossing rates are inversely proportional to the polymer length ($\sim1/N $)~\cite{Sebastian}. Because we are dealing with chains whose $R_{g}$ is comparable to $d_{0}$, our result (Fig.~\ref{fig:height}) lies between this prediction and that of Ref.~\cite{ParkSung}.

\begin{figure}
\includegraphics[width=8.5cm]{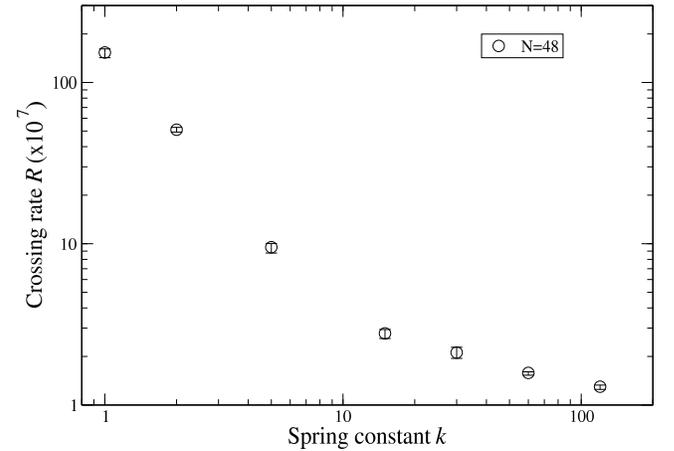}
\caption[0]{The flexible chain crossing rates as a function of spring constant $k$ with $N=48$. \label{fig:rate_spring48}}
\end{figure}

Figure~\ref{fig:rate_spring48} shows the crossing rates as a function of spring constant $k$ for chain length $N=48$. Smaller values of $k$ yield larger $\mathcal{R}$. This is because for small $k$, the chain can more easily extend at the barrier top, further reducing the activation energy \cite{SLee}. 

\subsection{Semiflexible polymer chain}

\begin{figure}
\includegraphics[width=8.5cm] {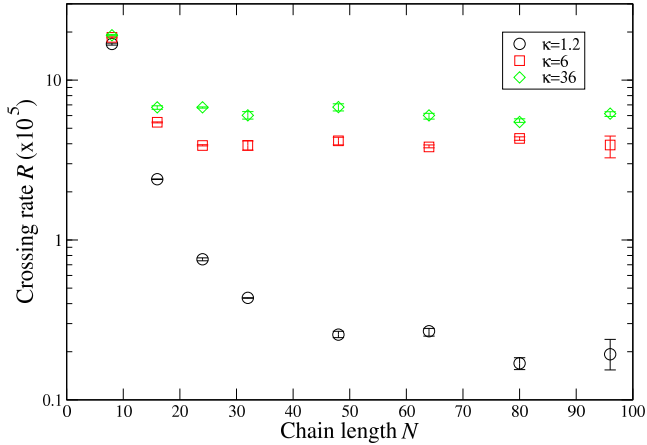}
\caption[0]{Semiflexible polymer barrier crossing rates with $k=30$. \label{fig:semiflexible_rates}}
\end{figure}

The bending stiffness characterizes prominently biopolymers; the persistence length is about $50$ nm for dsDNA~\cite{Taylor} and in the $10$ $\mu$m range for actin filaments~ \cite{kas1994direct, Howard}. We studied the semiflexible chain crossing over the barrier with a fixed stretching stiffness ($k=30$), and for three different values of bending stiffness $\kappa=1.2$, $6$ and $36$, corresponding to persistence lengths $2l_{0}$, $10l_{0}$ and $60l_{0}$, respectively. For different bending stiffnesses, Fig.~\ref{fig:height} and~\ref{fig:semiflexible_rates} show how the free energy barrier and the rates vary with chain length. For longer chains, the rates decrease toward constant values depending on the stiffness. 

The crossing rate for $\kappa=1.2$ decreases monotonically with $N$, similarly to flexible chains, but has a value higher than those of flexible chains. For $\kappa=6$ and $\kappa=36$, the crossing rates decrease with length and become nearly constant as $N$ increase above 24 and 16. As shown in Fig.~\ref{fig:height}, the free energy barrier height increases with $N$ until it reaches $N_{c}$, beyond which it decreases. For $\kappa=6$ and $\kappa=36$, the turnover chain length $N_{c}$ are 32 and 24, which are close to the lengths where the rates approach the plateaus. The match is not exact because although the exponential of $F_B/k_BT$ dominates the crossing rate, the rate prefactor still has a weak dependence on $N$.
  
\begin{figure}
\includegraphics[width=8.5cm]{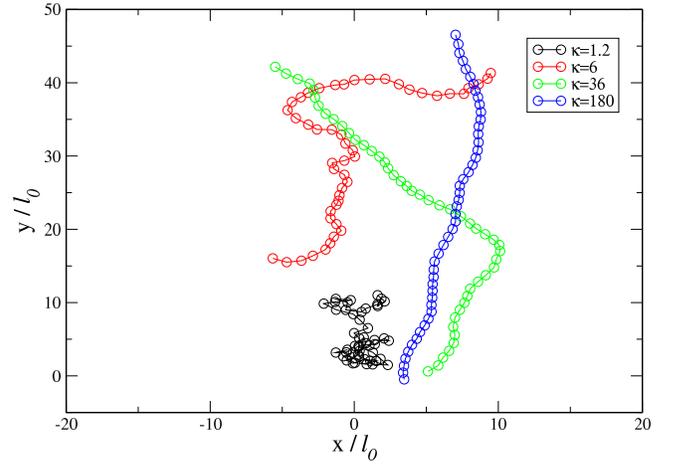}
\caption[0]{The chain configurations at the potential well for different bending stiffnesses and a fixed chain length $N=48$. The black markers correspond to $\kappa=1.2$, the red correspond to $\kappa=6$, the green correspond to $\kappa=36$, and the blue correspond to $\kappa=180$.
 \label{fig:equilibrium}}
\end{figure}

\begin{figure}
\includegraphics[width=8.5cm]{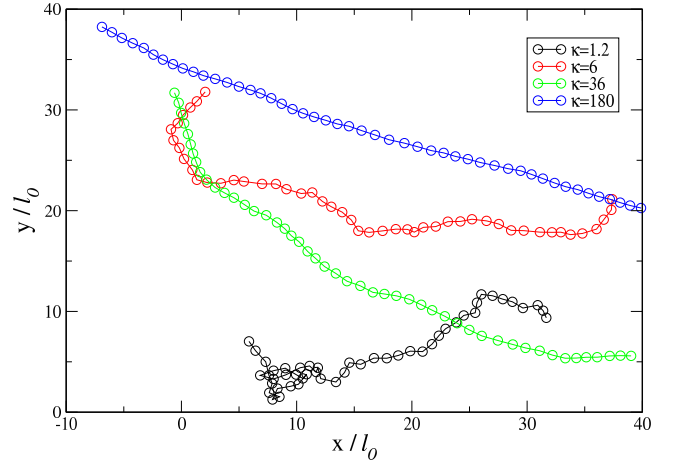}
\caption[0]{The chain configurations at the potential barrier (CM=16$l_{0}$) for different bending stiffnesses.  \label{fig:barrier}}
\end{figure}

The conformational behaviors that underlie these interesting results are shown in Figure~\ref{fig:equilibrium} and~\ref{fig:barrier} for $N=48$ and different bending stiffnesses. Within the potential well, stiffer chains with $\kappa \lesssim36$ become more extended and suffer a higher free energy. At the barrier top, the stiffer chain can more easily stretched along the $x$-axis and reduce the barrier height. Overall, stiffer chain can reduce the free energy barrier height ($F_{B}$) in a manner more pronounced for longer chain (see Figure~\ref{fig:height}). 

\begin{figure}
\includegraphics[width=8.5cm] {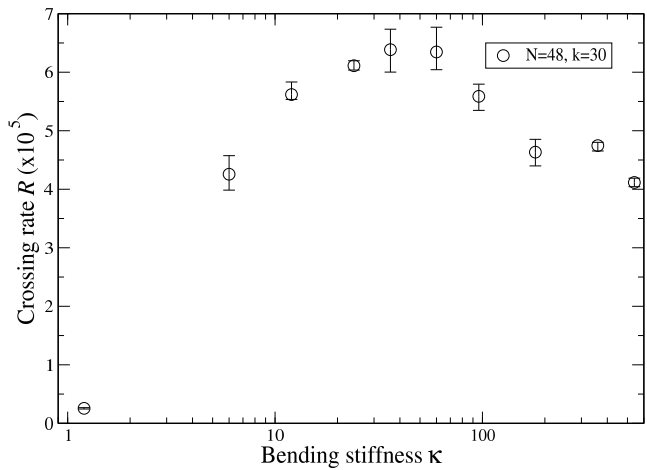}
\caption[0]{The crossing rates as a function of the bending stiffness for chain length $N=48$ and stretching stiffness $k=30$.\label{fig:kappa_rates}}
\end{figure}

Finally, Figure~\ref{fig:kappa_rates} shows the dependence of the crossing rate on bending stiffness $\kappa$ for $N=48$. The rate increases with $\kappa$ up to $\kappa\approx36$ (or persistence length$\approx60l_{0}$), above which it decreases. For $\kappa>36$, the chain, tending to align along the $y$ axis at the well, does not suffer the elevated free energy. With the CM located at the barrier top, this long and stiff chain is over the barrier and well bottom, so that the free energy is raised. Thus at an optimal value of $\kappa\approx36$, the free energy barrier height $F_{B}$ tends to be minimum, thereby yielding the maximum rate.

\subsection{Self-avoiding polymer chain}

\begin{figure}
\includegraphics[width=8.5cm] {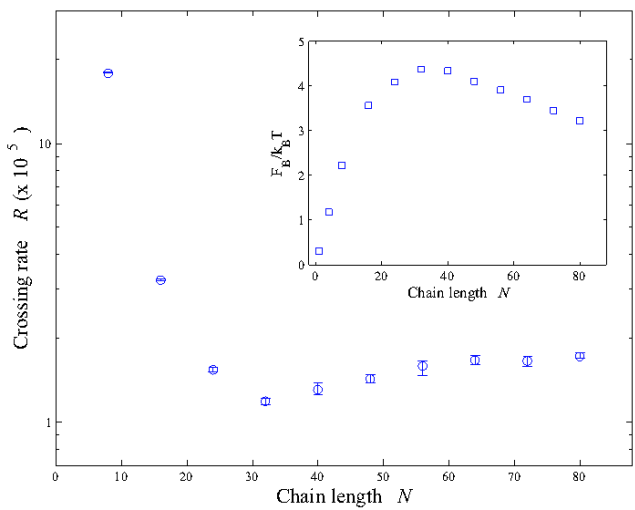}
\caption[0]{The crossing rates for the self-avoiding FENE chain as a function of chain length, with a minimum located at $N\approx 32$. Here, $k_\mathrm{F} = 15$, corresponding to the effective spring constant $k_\mathrm{eff}\approx 200$. The inset shows the free energy barrier height $F_B$ as a function of $N$. $F_B$ has a maximum at $N\approx 32$.  \label{fig:selfavoiding_rate}}
\end{figure}

To study the effect of excluded volume on the crossing rate for chains with $R_g$ comparable to the well size $d_0$, we also considered the self-avoiding FENE chain. In contrast to the flexible and semiflexible chains, we find that the crossing rate is a nonmonotonic function of chain length. The transition between a decreasing crossing rate and an increasing crossing rate occurs at $N\approx 32$, as shown in Figure~\ref{fig:selfavoiding_rate}. The minimum coincides with the maximum of the free energy barrier height $F_B$, as indicated in the inset of Figure~\ref{fig:selfavoiding_rate}. The behavior of $F_B$ and, consequently, the rate, is due to two competing factors. For short chains, the increase of $F_B$ is simply caused by the addition of particles to the chain. As the chain becomes longer, the excluded volume interactions cause the chain to swell, which forces the chain to occupy an increasingly wide region around both the well bottom and the barrier top. This causes the free energy barrier to decrease as a function of chain length after $N\approx 32$. The effect is similar to the case of semiflexible chain, but more pronounced.

The self-avoiding chain also exhibits a modest transition from a coiled state to a stretched state as it surpasses the barrier. However, the transition is weaker than for the flexible chain due to the geometry of the external potential. At the well, long chains tend to orient in the $y$ axis due to the excluded volume interaction and the external potential. Consequently, the conformation at the barrier top is less stretched along the $x$ axis than for the flexible chain, which starts the crossing in an almost isotropic configuration. For the self-avoiding chain, the effect of decreased free energy barrier due to chain swelling gives the dominant contribution to the enhanced crossing rate.

\section{Conclusions}
\label{sec:conclusions}

 We have studied the dynamics of polymer escape from a metastable Kramers potential using path integral hyperdynamics. Because the escape can be an extremely slow process, conventional simulations can demand enormous computing time. To speed up simulations, we have extended the path integral hyperdynamics (PIHD) method for polymer chains. We found that a constant bias force applied on each segment speeds up the rate without changing the chain configuration, allowing evaluation of the rate with a proper correction factor. 
 To demonstrate the efficiency of our simulation, we also computed conventional Langevin dynamics (LD) for some cases. For example, the semiflexible chain with $\kappa=1.2$ and $N=32$ case, PIHD takes about 1/30 of computing time via LD to get proper statistics.

  We considered flexible, semiflexible and self-avoiding chains with their radii of gyration $R_{g}$ smaller or comparable to the width of the potential well. We find that for a flexible chain, the crossing rate monotonically decreases with the chain length ($L$), but with much larger value compared to the chain's globular limit. For a semiflexible chain, the crossing rate becomes nearly constant as the chain becomes longer than a certain value. For a fixed chain length ($N=48$) the rate also shows nonmonotonic behavior as a function of the bending stiffness, exhibiting a maximum when the persistence lengh is about the contour length $L$. The enhancement of rates for the semiflexible chain over that of flexible chain can be interpreted as the reduction of the activation energy due to extended configuration of chain it takes in the potential well and at the barrier top. Finally, for a self-avoiding chain, the rate is a nonmonotonic function of chain length in contrast to the flexible and semiflexible chains. The reason for this is the excluded volume interaction, which causes the chain to swell and lowers the activation energy. For long chains this effect is more pronounced than for the flexible and semiflexbile chains, leading to rate increases with chain length. These findings suggest a possibility of polymer separation not only by its length but also by its bending stiffness.

\begin{acknowledgments}
This work was supported by NCRC at POSTECH and BK21 administrated by Korean Ministry of Education, Science and Technology, and in part by the Academy of Finland through its COMP Center of Excellence and TransPoly Consortium grants. T. Ikonen would also like to acknowledge the financial support of the Finnish Cultural Foundation and the Finnish Graduate School in Computational Sciences (FICS).
\end{acknowledgments}

\end{document}